\documentstyle[epsf,12pt]{article}

\begin{document}

\begin{titlepage}
\vskip 2cm
\begin{center}

{\Large \bf Further remarks on the isospin breaking in charmless 
semileptonic $B$ decays}
 \vskip 1cm

{\large G. L\'{o}pez Castro$^{a,b}$, J. H. Mu\~noz$^{b,c}$ and G. Toledo 
S\'anchez$^b$} \\

$^a$ {\em Institut de Physique Th\'eorique, Universit\'e catholique de
Louvain,}\\ {\em  B-1348 Louvain-la-Neuve, Belgium} \\

$^b$ {\em Departamento de F\'\i sica, Centro de Investigaci\'on y de  
Estudios} \\ {\em Avanzados del IPN, Apdo. Postal 14-740, 07000 M\'exico, 
D.F., M\'exico}

$^c$ {\em Departamento de F\'\i sica, Universidad del Tolima,}\\
{\em A.A . 546, Ibagu\'e, Colombia}

\end{center}

\vskip 2cm

\begin{abstract}
We consider the isospin breaking corrections to charmless semileptonic 
decays  of $B$ mesons. Both, the recently measured branching ratios of 
exclusive decays by the CLEO Collaboration  and the end-point region of 
the inclusive lepton spectrum in form factor models, can be affected by 
these corrections. Isospin corrections can affect the determination 
of $|V_{ub}|$ from exclusive semileptonic $B$ decays at a level 
comparable to present statistical uncertainties.  
\end{abstract}


\end{titlepage}%

\medskip

The first measurements of the exclusive charmless semileptonic decays of 
$B$ mesons have been reported recently by the CLEO collaboration 
\cite{cleo96}. A comparison between CLEO's results for the branching 
ratios of $B^0 \rightarrow \pi^- l^+ \nu$ and $B^0 \rightarrow 
\rho^- l^+ \nu$ and the theoretical expressions for their decay rates 
allows a determination of the $|V_{ub}|$ entry of the 
Cabibbo-Kobayashi-Maskawa mixing matrix. Actually, the value $|V_{ub}| = 
(3.3 \pm 0.2 ^{+0.3}_{-0.4} \pm 0.7) \times 10^{-3}$ has been estimated 
by combining the yields of five different channels of $B$ 
decays measured by CLEO 
and using four different theoretical models to describe the form factors 
of these exclusive decays.

  The small statistical uncertainty ($\approx 6\ \%$) quoted in CLEO's 
estimate of $|V_{ub}|$ is obtained by assuming the isospin symmetry 
relations 
\begin{eqnarray}
\Gamma(B^0 \rightarrow \pi^- l^+ \nu) &=& 2 \Gamma(B^+ \rightarrow 
\pi^0 l^+ \nu) \\
\Gamma(B^0 \rightarrow \rho^- l^+ \nu) &=& 2 \Gamma(B^+ \rightarrow 
\rho^0 l^+ \nu) \\ &=& 2 \Gamma(B^+ \rightarrow \omega l^+ \nu),
\end{eqnarray}
to combine the set of five channels in $B^+$ and $B^0$ decays into 
two independent measurements of $B(B^0 \rightarrow \pi^- l^+ \nu)$ and 
$B(B^0 \rightarrow \rho^- l^+ \nu)$.

  There are two reasons to consider the effects of isospin symmetry 
breaking in Eqs. (1) --(3). First, as we have shown in a previous paper 
\cite{prev}, the isospin breaking corrections due to $\rho^0-\omega$ 
mixing affect the relations (2)--(3) at the level of the statistical 
uncertainties reported for the $B^0 \rightarrow \rho^- l^+ \nu$ 
branching ratio \cite{cleo96}. Secondly, the precision for the branching 
ratios of exclusive charmless semileptonic $B$ decays would certainly be 
improved in forthcoming measurements at projected $B$ factories.

  In order to further emphasize the importance of isospin breaking 
effects let us mention that the correction to the $K^+ 
\rightarrow \pi^0 e^+ \nu$ decay ($K^+_{e3}$) due to the $\pi^0-\eta$ 
mixing affects its decay rate by around 3.4 \% \cite{lr}. The inclusion 
of this correction in semileptonic $K^+$ decays is important in order to 
achieve a determination of $|V_{us}|$ at the level of $\pm 1$ \% 
\cite{lr} by combining the semielectronic rates of $K^+$ and $K^0_L$ in a 
consistent way. As it was discussed by Leutwyler and Roos \cite{lr}, these 
isospin breaking corrections enter at first order in the charge of 
vector weak transitions without violating the Ademollo-Gatto theorem.

   Although the charmless semileptonic rates of $B$ mesons can not be 
measured with a similar precision as $K_{e3}$ decays, the several 
exclusive channels accessible to $B\rightarrow X_u l^+ \nu$ decays  
($X_u =\pi^-,\ \rho^-,\ \pi^0,\ \rho^0,\ \omega $) partially 
compensate the limited accuracy for the individual rates. In addition, the 
end-point region of the lepton spectrum in 
inclusive semileptonic $B$ decays is another source for the extraction of 
$|V_{ub}|$ \cite{pdg}. Those properties also suggest that is necessary 
to include isospin breaking corrections in semileptonic $B$ decays.

  In this {\it Brief Report} we first discuss the isospin 
breaking corrections to the lepton spectrum in inclusive 
charmless semileptonic $B$ decays as described by form factor models 
\cite{ff}. We 
also address some comments on the size expected for isospin breaking effects 
in the recent measurements of exclusive charmless $B$ decays as reported 
by CLEO \cite{cleo96}. 

  As is well known \cite{ff}, the inclusive lepton spectrum of $B$ decays 
in form factor models can be seen as a sum over exclusive channels. The 
end-point region of the lepton spectrum in semileptonic $B$ decays ($2.3 
\leq E_e \leq 2.6$ 
GeV, $E_e$ is the lepton energy) is expected to be dominated by a few 
exclusive modes ($B \rightarrow (\pi + \rho + \omega) l^+ \nu$). Most of 
the commonly used form factor models \cite{ff} explicitly assume isospin 
symmetry in their calculations. Neglecting resonances heavier than the
$\rho$ and $\omega$ masses, the lepton spectra in decays of neutral 
and charged $B$ mesons are thus given by
\begin{eqnarray}
\frac{d\Gamma^0(B^0)}{dE_e} &=& 
\sum_{X^-=\pi^-,\rho^-}\frac{d\Gamma^0(B^0 \rightarrow X^- l^+ \nu)}{dE_e} \\
\frac{d\Gamma^0(B^+)}{dE_e} &=&
\sum_{X^0=\pi^0,\rho^0,\omega}\frac{d\Gamma^0(B^+ \rightarrow X^0 l^+ 
\nu)}{dE_e} \nonumber \\
&=& \frac{d\Gamma^0(B^0)}{dE_e}-\frac{d\Gamma^0(B^+ \rightarrow \pi^0 l^+
\nu)}{dE_e},
\end{eqnarray}
where $d\Gamma^0(B \rightarrow X l \nu)/dE_e$ denote the lepton spectrum of 
each channel 
in the limit of exact isospin symmetry.

   After including isospin breaking corrections due to $\pi^0-\eta$ and 
$\rho^0-\omega$ mixing, the spectrum for the decay of $B^+$ mesons 
becomes (note that Eq. (4) remains unchanged):
\begin{equation}
\frac{d\Gamma(B^+)}{dE_e} = \left|1+\epsilon\right|^2 \frac{d\Gamma^0(B^+
\rightarrow \pi^0 l^+
\nu)}{dE_e}+\left|1+\epsilon'\right|^2\frac{d\Gamma^0(B^+ \rightarrow 
\rho^0 l^+
\nu)}{dE_e}+\left|1+\epsilon''\right|^2\frac{d\Gamma^0(B^+ \rightarrow 
\omega l^+ \nu)}{dE_e}
\end{equation}
where $\epsilon $ (see Leutwyler and Roos in \cite{lr}) and $ \epsilon',\  
\epsilon''$ \cite{prev} are given by
\begin{eqnarray}
\epsilon &=& \frac{3}{4} \frac{m_d -m_u}{m_s -\hat{m}} \approx 1.7 \ \times 
10^{-2}, \\
 \epsilon' &=& \frac{m_{\rho \omega}^2}{m_{\rho}^2-m_{\omega}^2+im_{\omega}
\Gamma_{\omega}} \approx 0.160 + 0.051 i, \\
\epsilon'' &=& -\ \frac{m_{\rho 
\omega}^2}{m_{\omega}^2-m_{\rho}^2+im_{\rho} \Gamma_{\rho}} \approx 0.006 
-0.031 i.
\end{eqnarray}
In the above expressions $m_q\ (q=u,d,s)$ denote current quark 
masses, $\hat{m} \equiv (m_u+m_d)/2$ and $m_{\rho \omega}^2 
=(-3.67 \pm 0.30) \times 10^{-3}$ GeV$^2$ \cite{blp}. The values for the 
resonance parameters of $\rho^0$ and $\omega$ mesons are 
taken from \cite{pdg}.
As discussed in ref. \cite{prev}, the value of $\epsilon'$ is very 
sensitive to the specific value used for the mass of the $\rho^0$.

 In order to illustrate these effects, 
  in Figure 1 we have plotted the lepton spectrum for decays of charged 
$B$ mesons by using the model of Isgur-Scora-Grinstein-Wise, 
Ref. \cite{ff}. The dashed plot  corresponds to the total spectrum in 
the limit of isospin symmetry, Eq. (5), and the corrected spectrum of Eq. 
(6) is represented by a solid line. As expected, the total spectrum 
corrected by 
isospin breaking is enhanced basically due to the second term in Eq. (6).

 Now we focus on the exclusive $B$ decays reported by CLEO \cite{cleo96}. 
 Isospin breaking corrections due to $\pi^0-\eta$ and $\rho^0-\omega$ 
mixings affect only the decays of charged $B$ mesons. After 
including these corrections, the relationships among the 
physical values of $B$ decays are changed to\footnote{Since vector 
mesons are resonances, a factorization approximation must be introduced 
to separate $B 
\rightarrow (\rho,\ \omega) l^+ \nu$ from the subsequent decay processes 
$\rho \rightarrow \pi \pi,\ \omega \rightarrow 3\pi$. In fact, our 
results on isospin breaking can also be derived by using the mixed 
propagator matrix of the $\rho^I-\omega^I$ unstable system ($I$ is for 
isospin 
eigenstates) to describe the full $B \rightarrow (2\pi,\ 3\pi) l^+ \nu$ 
decay processes.} \cite{prev}: 
\begin{eqnarray}
\Gamma(B^0 \rightarrow \pi^- l^+ \nu) &=& 2 \Gamma(B^+ \rightarrow
\pi^0 l^+ \nu)/|1+\epsilon |^2 \\
\Gamma(B^0 \rightarrow \rho^- l^+ \nu) &=& 2 \Gamma(B^+ \rightarrow
\rho^0 l^+ \nu)/|1+\epsilon' |^2 \\ &=& 2 \Gamma(B^+ \rightarrow \omega l^+ 
\nu)/|1+\epsilon'' |^2, 
\end{eqnarray}
to be compared to Eqs. (1)--(3).
  According to Eqs. (7)--(9), the effect of isospin breaking on the $B^+ 
\rightarrow \pi^0 l^+ \nu$ and $\omega l^+ \nu$ decay modes is negligible 
compared to form factors and present experimental uncertainties. The most 
important effect would be present in $B^+ \rightarrow \rho^0 l^+ \nu$.

   In practice, it turns out impossible to compute the effects of Eqs. 
(10)--(11) in the extraction of $B(B^0 \rightarrow \rho^- l^+ \nu)$ and 
the value of $|V_{ub}|$ from the numbers quoted in Ref. \cite{cleo96}.
 Actually, the relative normalizations imposed by isospin constraints,
(1)--(3) or (10)--(12), must be implemented in the simultaneous fit to the
five semileptonic channels. In addition, the yields depend on the
kinematical region included in the analysis because, for instance, the
signal and the cross feed among different modes smear outside the region
considered in \cite{cleo96}.
 
  The CLEO Collaboration has redone the analysis of Ref. \cite{cleo96} by 
implementing Eqs. (11)--(12) \cite{gibbons}. Using the model of 
Ref. \cite{isgwII} as an example, CLEO's reanalysis 
finds that $B(B^0 \rightarrow \rho^- l^+ \nu)$ shifts by --8.8 \% in 
the combined fit to the $\rho^-/\rho^0/\omega$ channels \cite{gibbons}. 
Therefore, the corresponding value of $|V_{ub}|$ shifts only by --4.4 \% 
from the value obtained when isospin symmetry is assumed \cite{cleo96}.
If the $B \rightarrow \pi l \nu$ modes are also considered, the value of
$|V_{ub}|$ is further reduced to --3.3 \%.

   In summary, even though isospin breaking corrections to the exclusive 
channels reported in Ref. \cite{cleo96} are smaller than 
present total uncertainties, they are important effects to be included in 
forthcoming improved measurements of inclusive and exclusive 
charmless semileptonic decays of $B$ mesons. To conclude, let us mention 
that similar corrections due to $\rho^0-\omega$ mixing must be applied 
when combining different decay channels to quote an average for the 
flavor changing radiative decays  $B \rightarrow \rho + \gamma$.

\medskip

{\bf Acknowledgements}

 We are grateful to H. Castilla for useful conversations. We are 
specially indebted to L. K. Gibbons for communication of CLEO's 
reanalysis of the data of Ref. \cite{cleo96} including isospin breaking 
effects. The authors would like to  acknowledge the financial support 
from Colciencias (J.H.M.) and Conacyt (G.L.C. and G.T.S.).

\newpage

\begin{figure}[t]
\leavevmode
\par
\begin{center}
\mbox{\epsfxsize=14.cm\epsfysize=16.cm\epsffile{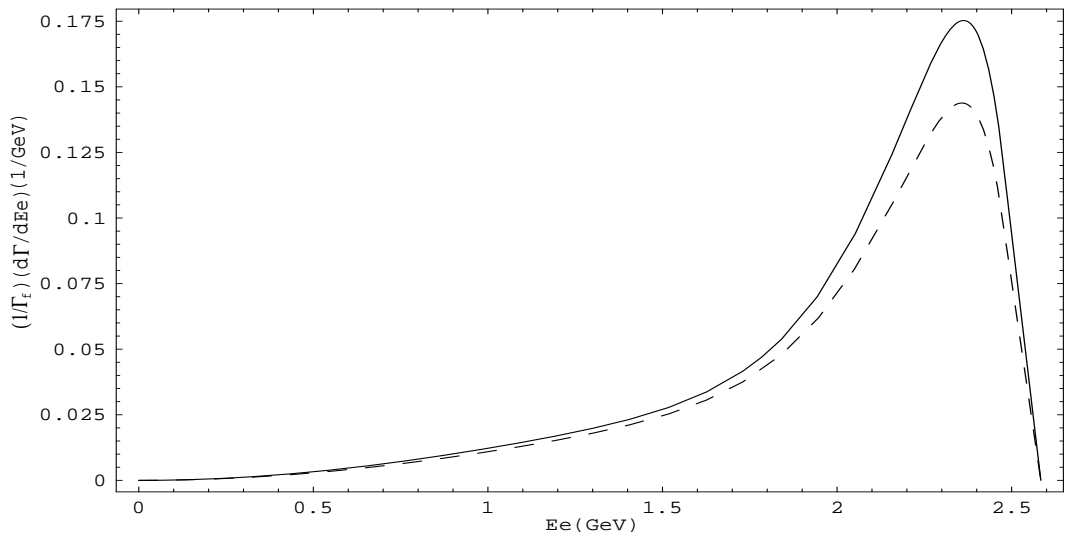}}
\end{center}
\caption{ Inclusive lepton spectrum in charmless semileptonic $B^+$ decays
with (solid line) and without (dashed) isospin breaking corrections.
These  spectra are normalized to the tree level width $\Gamma_f$ for $b
\rightarrow ul^- \nu$. } 
\end{figure}


\medskip

\begin{thebibliography}{9}
\bibitem{cleo96} CLEO Collaboration, J. P. Alexander {\it et al.}, Phys. 
Rev. Lett. {\bf 77}, 5000 (1996).

\bibitem{prev} J. L. D\'\i az-Cruz, G. L\'opez Castro and J. H. Mu\~noz, 
Phys. Rev. {\bf D54}, 2388 (1996).

\bibitem{lr} H. Leutwyler and M. Roos, Z. Phys. {\bf C25}, 91 (1984); G. 
L\'opez Castro and J. Pestieau, Mod. Phys. Lett. {\bf A4}, 2237 (1989); G. 
L\'opez Castro and G. Ordaz Hern\'andez, Mod. Phys. Lett. {\bf A5}, 755 
(1990).

\bibitem{pdg} Particle Data Group, R. M. Barnett {\it et al.},
 Phys. Rev. {\bf D54} (Part I), 1 (1996).

\bibitem{ff} N. Isgur, D. Scora, B. Grinstein and M. B. Wise, Phys. Rev. 
{\bf D39}, 799 (1989); M. Wirbel, B. Stech and M. Bauer, Z. Phys. {\bf 
C29}, 637 (1985); J. G. K\"orner and G. A. Schuler, Z. Phys. {\bf C38}, 
511 (1988).

\bibitem{blp} A. Bernicha, G. L\'opez Castro and J. Pestieau, Phys. Rev. 
{\bf D50}, 4454 (1994). 

\bibitem{gibbons} L. K. Gibbons, private communication.

\bibitem{isgwII} N. Isgur and D. Scora, Phys. Rev. {\bf D52}, 2783 (1995). 

\end{thebibliography}
\end{document}